\documentclass[a4paper,fleqn,usenatbib,useAMS,referee]{mnrasDraft}

\usepackage{graphicx}	
\usepackage{amsmath}	
\usepackage{amssymb}	
\usepackage{multicol}        
\usepackage{bm}		
\usepackage{pdflscape}	

\newcommand{\kms}{\,km\,s$^{-1}$} 

\usepackage[T1]{fontenc}
\usepackage{ae,aecompl}

\usepackage{txfonts}

\title[SN~2015bh in NGC 2770]{Photometry and spectroscopy of SN~2015bh in
the galaxy NGC~2770}

\author[V. P. Goranskij et al.]{
V.~P.~Goranskij,$^{1}$
\thanks{Contact e-mail: \href{mailto:goray@sai.msu.ruk}{goray@sai.msu.ru}}
~E.~A.~Barsukova,$^{2}$
~A.~F.~Valeev,$^{2,3}$
~D.~Yu.~Tsvetkov,$^{1}$
\newauthor 
~I.~M.~Volkov,$^{1}$
~V.~G.~Metlov,$^{1}$ 
~A.~V.~Zharova$^{1}$
\\
 $^{1}${Sternberg Astronomical Institute,  Lomonosov Moscow State University, Universitetsky Prospect, 13, Moscow 119899 Russia}\\
 $^{2}${Special Astrophysical Observatory of the Russian Academy of Sciences, Nizhnij Arkhyz, Karachay-Cherkessia  369167 Russia}\\
 $^{3}${Kazan Federal University, Kremlevskaya street 18, Kazan, Tatarstan 420008 Russia}
}

\date{\today}

\pubyear{2016}

\begin{document}
\label{firstpage}
\pagerange{\pageref{firstpage}--\pageref{lastpage}}
\maketitle

\begin{abstract}
We present medium-resolution spectroscopy and multicolor photometry for
the optical transient PSN J09093496+3307204 (SN~2015bh) in the galaxy NGC~2770,
which has transferred into the supernova phase. The observations were carried
out between February 2015 to May 2016.
Both at the phase of the SN impostor (2015a) and at the supernova phase
(2015b), besides Balmer emissions, the strong Fe~II emissions are seen
in the spectrum; so, these spectra resemble those of Williams Fe~II
type classical novae. The star is located near the edge of a dark nebula and 
notably absorbed (A$_V$ = 1\fm14 $\pm$0\fm15). Taking into account this
absorption, we determined maximum absolute magnitudes of 
$M_V$ = --15\fm0 $\pm$0\fm3 at the 2015a phase and of
$M_V$ = --18\fm14 $\pm$0\fm30 at the 2015b phase. The light curve at the 2015b
phase is similar to those of SN IIL. The supernova progenitor is a luminous blue
variable (LBV) star with the powerful H$_\alpha$ emission. We considered several
hypotheses of supernovae explosions following optical transients related
with LBV. The hypothesis of core collapse of an evolved massive star interrupting
the process of its merging with massive companion in a binary system (a failed
luminous red nova) was chosen as the preferable one for this event. 

\end{abstract}



\section{Introduction}
The optical transient PSN J09093496+3307204 was discovered in the galaxy
NGC 2770 on February 7, 2015 in the Catalina Real-Time Transient Survey, 
and on February 8, 2015 by S. Hoverton in the process of observations under
the program SNhunt\footnote{http://www.rochesterastronomy.org/sn2015/snhunt275.html}. 
The transient is also known as SN 2015bh and SNhunt275. 
In the spectrum obtained in Asiago Observatory on February 9, a broad emission
H$_\alpha$ (FWHM $\sim$ 6800 \kms) with a narrow line with P Cygni
profile on top was visible \citep{ERea15}. The absorption component of the profile
showed the expansion velocity $\sim$950 \kms.
In 2008 -- 2009 HST images, a faint star was detected coinciding in coordinates 
with the transient and variable within 21\fm5 -- 22\fm8 in the red filter F606W.
Its spectrum and chaotic variability resembled supernova impostors 
2000ch and 2009ip, before the explosion of latter as a supernova in
June 2012 \citep{ERea15}. The definition of "supernova impostor" was introduced
by \citet{VDPKea00} to identify the explosions of LBV stars in which stars
survived and being observed several years after outbursts in contrast to SN~IIn,
which were destroyed at the core collapse.  The eruption of $\eta$ Car in 1844 -- 1850
can be an analog of a SN impostor. So, the transient in NGC 2770 was classified
as a SN impostor or an LBV explosion. 

Photometry and spectroscopy with the 10.4 m GTC telescope at Canarias within
the time interval between March 27 and April 14, 2015 revealed a considerable
increase of star brightness by 1\fm2, and the object reached the absolute
magnitude --14\fm2 in the SDSS $r$ band \citep{UPLea15}. In the spectrum
taken on April 14, the H$_\alpha$ emission, other Balmer lines, the distinct
emissions Fe~II in the range $\lambda$ 4950--5400 \AA\ and the Na~I/He~I blend
were predominating.  The H$_\alpha$ line profile was asymmetric with a red-side
wing extending up to $\sim$5000 \kms, and a steeper blue-side decline.  
But the P~Cyg type profile in H$_\alpha$ line was already not observable. 

On May 16, 2016, \citet{UPTea15} informed about a sharp increase of the transient
brightness by 2$^m$. The absolute magnitude in the $R$ band reached \hbox{--16\fm4.}
In new spectra of May 16 from the Keck~I telescope, the star displayed a hot
continuum with Balmer, He~II and He~I emission lines \citep{DBLea15}.
H$_\alpha$ remained asymmetric with the half-width FWHM $\sim$1200 \kms.
It became clear that it was a new explosion of the impostor, and the events were 
developing according to the scenario of SN 2009ip \citep{MSF13,MMSea14,GSVea14,PCIea13}.
The observations continued by \citet{CTLea15,VSV15} and \citet{RA15} confirmed
the assumptions that the transient passed to the SN~II phase. By analogy with
other papers, we designate the first outburst or the impostor phase as 2015a and
the second brightening or the SN phase as 2015b. 

The host galaxy of SN~2015bh, NGC~2770 is located at a distance of d =
29.70 $\pm$3.4 (distance module $M-m$ = 32\fm35 $\pm$0.24); its radial
velocity is $v_r$ = 1947 \kms\ (redshift z = 0.006494); the Galaxy
extinction in this direction is $A_V$ = 0\fm062 (NED).
Before the 2015 event, three SNe of the rare Ib type exploded in the galaxy
NGC~2770 during a short period of 10 years, that is why the galaxy was called
"a factory of type Ib supernovae"\ \citep{TMLea09}. These are SNe 1999eh, 2007uy
and 2008D. The latter one is interesting by the fact that it was related to
an X-ray transient. \citet{TMLea09} show in their Fig.~1 the galaxy image
assembled from three VLT frames of March 16, 2008, including one with the filter
centered on 6604~\AA\ at $z$ = 0.007 and the width 64~\AA, and another one in the
filter H$_\alpha$ at $z$ = 0 with the center at 6563~\AA\ and the width 61~\AA. 
Besides the two SNe of 2007 and 2008, one can see there a SN~2015bh progenitor,  
which is distinguished by its excess of emission in the H$_\alpha$ line.
This is one of the brightest stars of the galaxy. 

\section{PHOTOMETRY}

Photometry in the Johnson $UBV$ and Cousins $RI$ bands was taken with the
focal reducer SCORPIO at the BTA \citep{Afanas05} in the process of spectral
observations. Additionally, CCD photometers were used at the 1-m telescope of SAO
RAS, at the 70-cm telescope of SAI MSU in Moscow, at the 60-cm reflector and the
50-cm Maksutov meniscus telescope of the Crimean Observational Station of MSU,
and at the 1-m and 60-cm telescopes of the Simeiz Observatory in Crimea.
Our observations were performed in the period from February 24, 2015 to May 30,
2016 with a break from June 16 to September 10, 2015, when the object was not
observed due to conjunction with the Sun. 
The first images after the break suitable for measurement were obtained with the
1-m telescope of SAO RAS in the $R_C$ and $I_C$ filters on September 11, 2015, 
against the dawn background. The standard stars from \citet{MBCea09}
near NGC~2770 in the Johnson $UBV$ and Cousins $RI$ system
were used as comparison stars. Altogether we have 43 observations, both multicolor
ones and those taken in individual filters. They are listed in Table~1. The full
collection of observations of SN~2015bh including published observations of
other authors and supplemented by new data is available in 
Internet\footnote{http://jet.sao.ru/$\sim$goray/psn0909.ne3}$^,$\footnote{http://vgoranskij.net/psn0909.ne3}. 
The light and color curves can be examined with a Java-compatible 
browser\footnote{http://jet.sao.ru/$\sim$goray/psn0909.htm}$^,$\footnote{http://vgoranskij.net/psn0909.htm}.

\begin{figure*}
\includegraphics[scale=1.3]{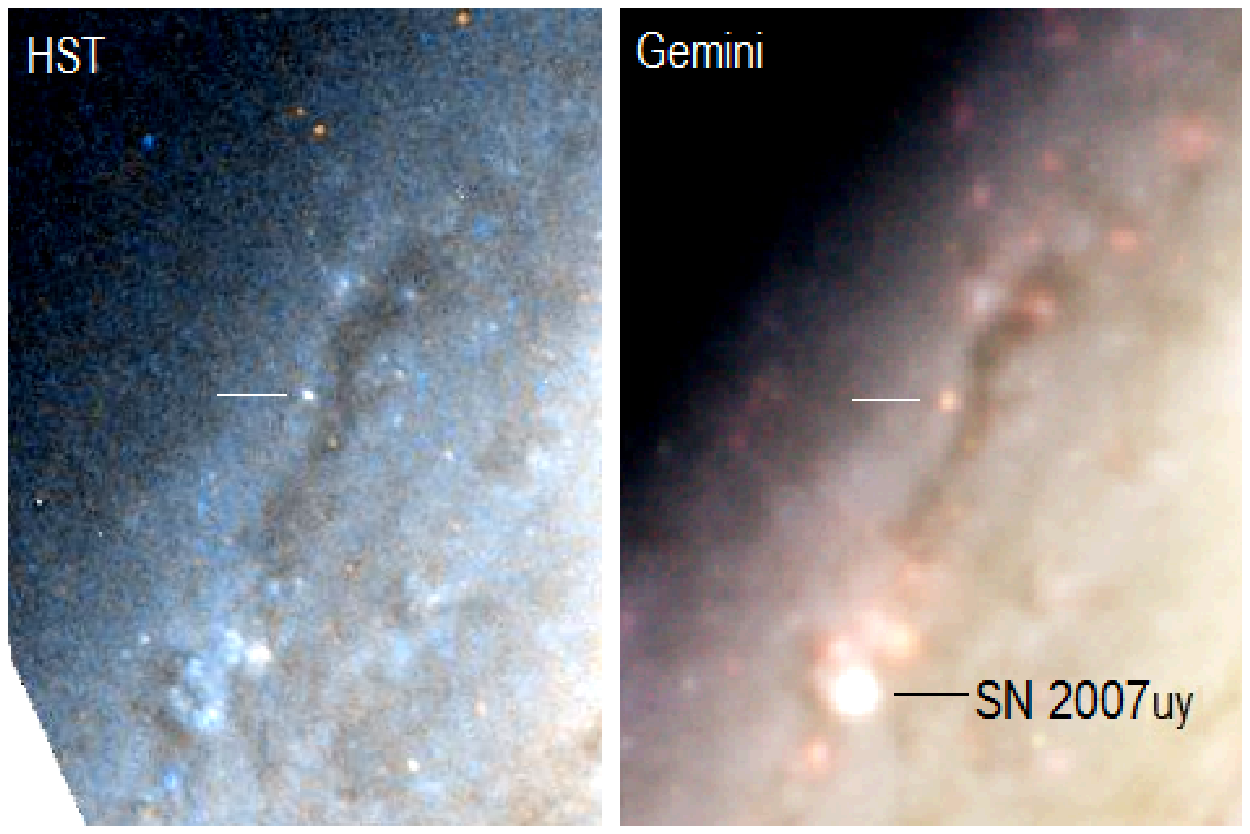}
\caption{
Fragments of high angular resolution images of the region
40$\arcsec\times$60$\arcsec$ around SN 2015bh the galaxy NGC~2770.
Left: the image from the Space Hubble Telescope of December, 19, 2008;
right: an image from the "Gemini"\ telescope of March 6, 2008. The progenitor
is marked by the horizontal line. 
}
\label{figure1}
\end{figure*}

The accuracy of measurements is given in Table~1 in brackets after each
observation in units of the last significant digit. The Table comment contains
information  on telescopes and detectors used in observations.
There are small systematic differences between sets of our observations 
obtained with different devices. These differences were determined from
simultaneous measurements and compensated for corrections. After the conjunction
with the Sun when the star brightness was faint, we were accumulating
signal from the star during a long time, 1300 -- 6000 s, to achieve a high
signal-to-noise ratio. 
The main factor of the errors during long accumulation was an inhomogeneous
surrounding background of the galaxy, because the star is located in a spiral
arm near a dark dust nebula. Fig.~\ref{figure1} presents fragments of images 
with the highest angular resolution taken with the Hubble Space Telescope and 
with the ground-based telescope Gemini (a fragment of the color image
by S.~Rankin in Flickr.com). The progenitor, an LBV star is identified in
the image. The HST images have been already analyzed by \citet{ERea15}.
We have measured the color image of Gemini of March 6, 2008 
composed from components in the SDSS $gri$ filters and H$_\alpha$ filter by
decomposing it into RGB components. 
The faintest stars from the SDSS database in the vicinity of the galaxy NGC 2770 
were used as comparison stars. At that moment the star brightness was
$g = 25\fm4 \pm0\fm15$ and $r = 25\fm0 \pm0\fm2$ in the system AB95. 
These values correspond to $V$ = 25\fm2. In the filters $i + H_\alpha$ 
the star was much brighter, 23\fm3 $\pm$0\fm2.

\begin{figure*}
\includegraphics[scale=0.7]{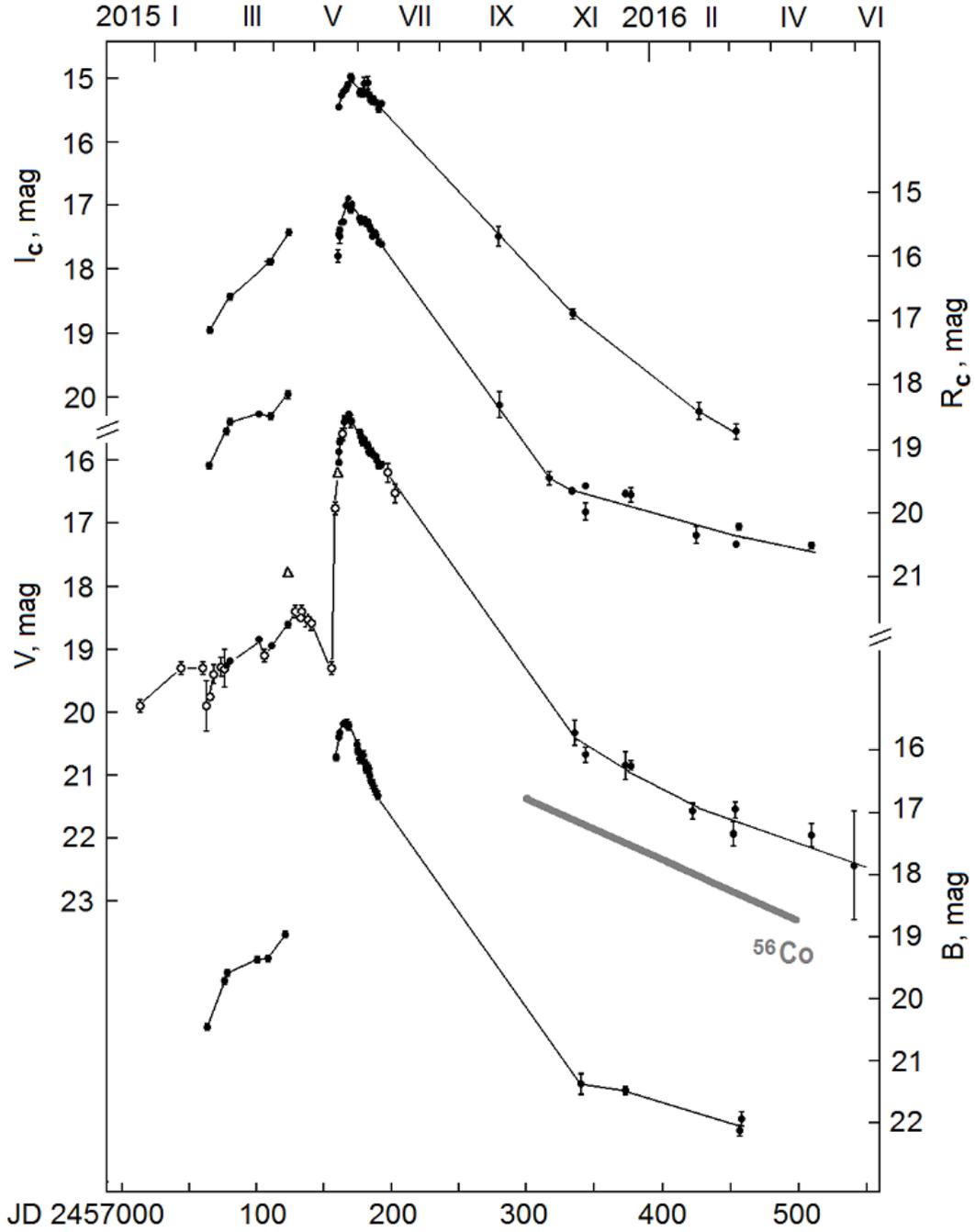}
\caption{Light curves of SN 2015bh in 2015--2016 in $B,V,R,I$ bands 
(from bottom to top). The points are multicolor CCD observations including
our ones and those ones cited in the text. The circles are observations from the
collection by S.~Hoverton obtained mainly with CCD without a filter. 
The triangles are observations by Zhijang Xu and Xing Gao. The gray line shows
the average rate of brightness decay of SN I and II caused by radioactive decay
of $^{56}$Co. 
}
\label{figure2}
\end{figure*}

The $BVR_CI_C$ light curves are shown in Fig.~\ref{figure2} (from bottom to top).
Besides our observations, here the data published by \citet{ERea15,UPLea15} 
(converted from SDSS values to the $UBVR_CI_C$ system) and by 
\citet{CTLea15,VSV15} are collected. These observations and our ones are
shown by points. Observations from a collection of 
S.~Hoverton\footnote{https://www.flickr.com/photos/watchingthesky/17162602585/},
A.~Cason and 
R.~Arbour\footnote{http://www.rochesterastronomy.org/sn2015/snhunt275.html} 
are plotted on the $V$ band light curve by circles. These observations were
made with different CCDs and mainly without filters, but they are consistent
with our observations in the $V$ filter rather well. Triangles denote observations
by Zhijian Xu and Xing Gao from the Hoverton's collection.
The first observation (2015 April 7.74~UT, mag = 17.8, no filter) is
considerably (by 0\fm8) brighter than the mean light curve and the observation 
by \citet{UPLea15} obtained at nearly the same time. It is possible that they
registered a short flare. 
Separate observations by G.~Locatelli (Flickr.com) and G.~Masi (CBAT) 
are also plotted with circles. 

The phase 2015a is covered by observations from December 21, 2014 to April 25, 2015,
or within JD 2457013 -- 2457138.
At that time photometry shows a gradual rise of brightness with an average rate 
of 0\fm015 per day (0\fm75 during 50 days). 
The observed brightness fluctuations in amplitude were of several tenths of a
magnitude, there were
a short brightness weakening by 0\fm9 near JD 2457066 and brightness decline
by about 1$^m$ before the rise to the main maximum. 
At the phase 2015a, the star achieved the level 18\fm4 (CCD without filter). 
The decline of brightness before the rise to the main maximum (JD 2457153, 19\fm3)
registered by Hoverton is a very important observation for the understanding  
the nature of 2015bh, but it should be noted that this observation is single. 
The SN brightness rise (the beginning of the phase 2015b) has been registered 
starting from May 13, 2015 (JD 2457156). The brightness maximum of the phase 2015b
was on May 22, 2015, at the moment JD 2457165 $\pm1^d$.
Maximum magnitudes at the phase 2015b were as follows: $U$ = 14\fm75,
$B$ = 15\fm57, $V$ = 15\fm35, $R_C$ = 15.21, $I_C$ = 15.10 measured
with the SAO 1 m telescope, and $B$ = 15\fm54, $V$ = 15\fm35, $R_C$ = 15\fm18
measured with the 60 cm telescope of Simeiz Observatory. In the SN light
curve (2015b), we allocate the initial linear decay of brightness after maximum
of May 22 -- October 17, 2015 (JD 2457165--2457313) and the secondary
decay after October 17, 2015 (JD > 2457313). An abrupt transition in the
brightness decay rate from the initial one to the secondary one is seen in all
bands except $I_C$.
In the initial decay in the first 50 days, the rates of brightness decline in the
$BVR_CI_C$ filters were, respectively, 2\fm0, 1\fm5, 1\fm3 and 1\fm0.\ A sharp peak in
the maximum and a linear brightness decay in the magnitude scale continued during
$\sim$150 days after the maximum are typical for light curves of SN~IIL.
A considerable part of this period, the star was inaccessible for observation.
However, September observations in the $R_C$ and $I_C$ filters 
carried out in SAO RAS right after appearance of the object from behind the Sun
confirm the linear character of the initial decay. The mean light curve of
SNe~IIL\footnote{http://burro.cwru.edu/academics/Astr221/LifeCycle/observingSN.html}
shows that the brightness decline rate changes within the range of 2\fm2 -- 3\fm0
during 50 days in the $B$ filter.
SN~2015bh demonstrates a lower brightness decay rate (2\fm0) in comparison with
this summary light curve. Yet, when comparing the decay rate in the $V$ filter
with three separate SNe~IIL 1980K (1\fm9),
1998S (1\fm5)\footnote{http://dau.itep.ru/sn/lc} and 2008if (1\fm4) \citep{Aea14}
the light curve of SN 2015bh looks typical for SNe~IIL.
Then, in the secondary decay at the end of October 2015, the rate of brightness
decline reduced considerably, especially in the filter $R_C$.
The slowest brightness decay rate in $R_C$ is obviously related to a gradual
increase of contribution of the H$_\alpha$ emission along with the weakening of
continuum in the long-wave range. In Fig.~\ref{figure2}, the heavy gray line shows
the secondary light decay rate due to radioactive decay of the $^{56}$Co isotope (as
in SNe~I and in several SNe~II). The average slope in the 200-day secondary
brightness decay in the $V$ filter (where the contribution of emission lines
is minimum) was 0\fm0092 per day (0\fm46 during 50 days), what corresponds to
the hypothesis of the $^{56}$Co decay rate and confirms the assumption that
the event 2015b was a SN. 

\begin{figure*}
\includegraphics[scale=0.75]{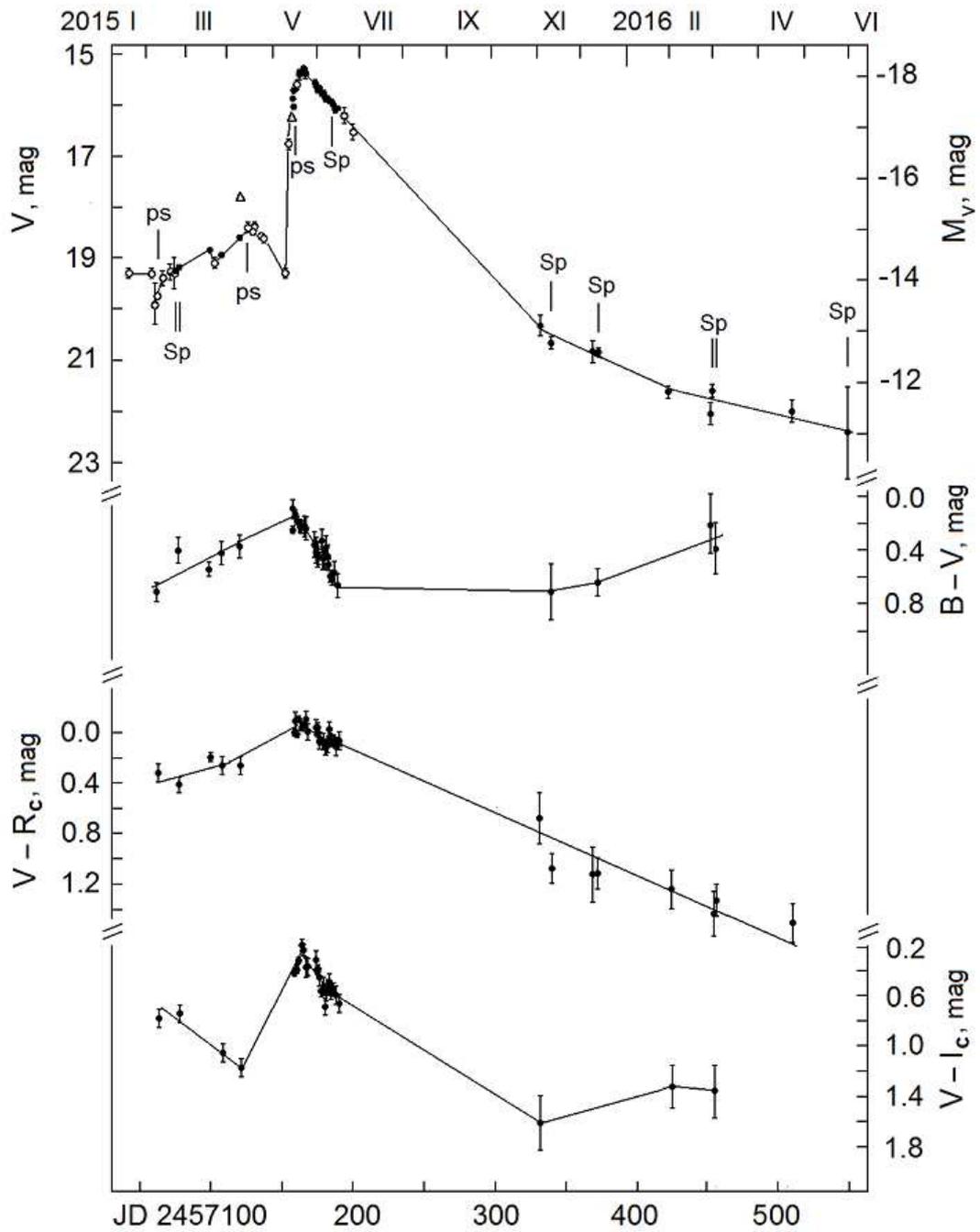}
\caption{Light curve of SN~2015bh in the $V$ band in the scale of absolute
magnitudes and $B - V$, $V - R_C$ and $V - I_C$ color index curves built 
from observations of 2015 -- 2016. The moments of spectral observations taken
in this search are marked by a label {\it Sp}, and other spectral observations 
published in Internet, or if their description are published, are marked by {\it ps}. 
}
\label{figure3}
\end{figure*}

The color-index $B-V$, $V-R_C$ and $V-I_C$ curves are presented in 
Fig.~\ref{figure3} together with the $V$  light curve in the scale 
of absolute $V$ magnitudes (right). When plotting the color curves, we used not
only simultaneous observations obtained in separate nights, but also the
observations in consecutive nights in the interval 1--2 days. When the star was
faint, and the light accumulation time amounted to 2 hours, it was impossible
to do that for all filters during one night. On the other hand, in the secondary
decay, the changes of the object brightness during 2 days were much lower than
measurement errors, so, the color indices are quite realistic. At the phase 2015a,
$B-V$ and $V-R_C$ color indices were gradually reducing, what reflected
increasing star's temperature.
However, the $V-I_C$ trend was opposite. At the explosion of SN
at the phase 2015b, the star was the hottest in the rising branch at the moment
JD 2457159, i.e. 6 days before maximum, with the color indices $U-B$ = --0\fm85,
$B-V$ = +0\fm08 (this is data of SWIFT/UVOT in the system $ubv$ AB
\citep{CTLea15}, converted into the Vega $UBV$ system). This fact is confirmed
by spectral observations  with the Keck telescope \citep{DBLea15}, when the hot
continuum and the He~II 4686 \AA\ emission were observed.
Taking into consideration the fact that
at that time $UBV$ bands were predominated by the hot continuum, and 
the contribution of emissions was small (it is expected to be <3\%) we determined 
the color excess $E(B-V)$ = 0\fm37 $\pm$0\fm05 and extinction 
$A_V$ = 1\fm14 $\pm$0\fm15. This estimate of interstellar extinction includes
both the extinction in our Galaxy and that in the galaxy NGC 2770 together with
the circumstellar environment. With due regard to such extinction along the line of
sight, the absolute visual magnitude of SN in the brightness maximum was 
$M_V$ = --18\fm14 $\pm$0\fm30, and in the impostor stage it achieved 
$M_V$ = --15\fm0$\pm$0\fm3 (or even $M_V$ = --15\fm7, if relying on the
observation by Zhijian Xu and Xing Gao discussed above).
Thus, it follows from observations that the event 2015b with its light curve shape
and absolute magnitude in maximum can be interpreted as a real SN.  

From measurements of minimum brightness on March 6, 2008, with the telescope
Gemini $V$ = 25\fm2 $\pm$0\fm3 with allowance made for extinction, the
absolute magnitude of the SN~2015bh progenitor was M$_V$ = --8\fm3 $\pm$0\fm3.
This is close to the \citet{HD79} limit for the most massive stars
and confirms classification of the progenitor as an LBV. Stars of so high luminosity
are not stable, they lose their mass and form the circumstellar medium rich with
gas and dust. 

Starting from the first moment of registration of the SN on the ascending
branch of the 2015b outburst, a gradual reddening in all the color indices 
continued until the break in observations caused by conjunction with the Sun. 
It is obvious that this reddening occurred due to expansion of the ejected 
optically thick gas shell and decreasing of surface temperature of this shell. 
However, after the break in observations, the behavior of color indices changed.
$B-V$ and $V-I_C$ indices became bluer and revealed opposite trends in their
changes, whereas the reddening of $V-R_C$ continued with the same rate.
We explain that behavior by transition of the shell in the optically thin state, what
leads to the fact that the radiation passes from hot inner layers of the shell. 
However, in the $V-R_C$ index, the contribution of the H$_\alpha$ emission 
increases relative to the hot continuum, and the increase of this index
continues. The color index $V-R_C$ increased by $1\fm4$ during 9 months after
the brightness maximum. Similar changes of color indices occur in some classical
novae, and they are related to analogous changes in the structure of ejected
shells.

\section{SPECTROSCOPY}

Moderate-resolution spectral observations were carried out with the Russian 
6 m telescope BTA and the focal reducer SCORPIO \citep{Afanas05} 
during 8 nights in the period between February 23, 2015 and May 30, 2016. 
The observations were obtained 88 and 87 days before the SN brightness maximum,
in the initial brightness decay 22 days after this maximum, and in the secondary
decay, in the time range between 176 to 374 days after the maximum. Table~2
contains main information on the spectra: date, and Julian date, time in days
counted from the brightness maximum, full exposure in seconds, spectral range,
spectral resolution, grism, heliocentric corrections to radial velocities,
and signal-to-noise ratio in continuum in the middle of the spectral range. 
Naturally, the signal-to-noise ratio in the profiles of spectral lines is much
higher. 
The spectra were processed in OS Linux using the ESO MIDAS package and the LONG
context (for the long-slit spectra). In the red wavelength range of
$\lambda$ > 6800 \AA, the spectra are distorted by interference pattern (fringes). 
Since the signal-to-noise ratio in spectra obtained in November 2015 and later
is too small (2--4), we smoothed them by the moving-average method
with the averaging interval 3.5 -- 15 \AA, which is equal to actual spectral
resolution of each spectrum. To convert the spectra into energy units, we used
spectrophotometric standards by \citet{Oke90} and simultaneous photometric
observations. The spectra are available in digital form in 
Internet\footnote{http://jet.sao.ru/$\sim$bars/spectra/psn0909/}.
The moments of our spectral observations are plotted in the light curve 
(Fig.~\ref{figure3}, {\it Sp}). Spectral observations published by other authors are
also marked in the Figure with ({\it ps}); these spectra or their descriptions are
available in Internet. 

\begin{figure*}
\includegraphics[scale=0.7]{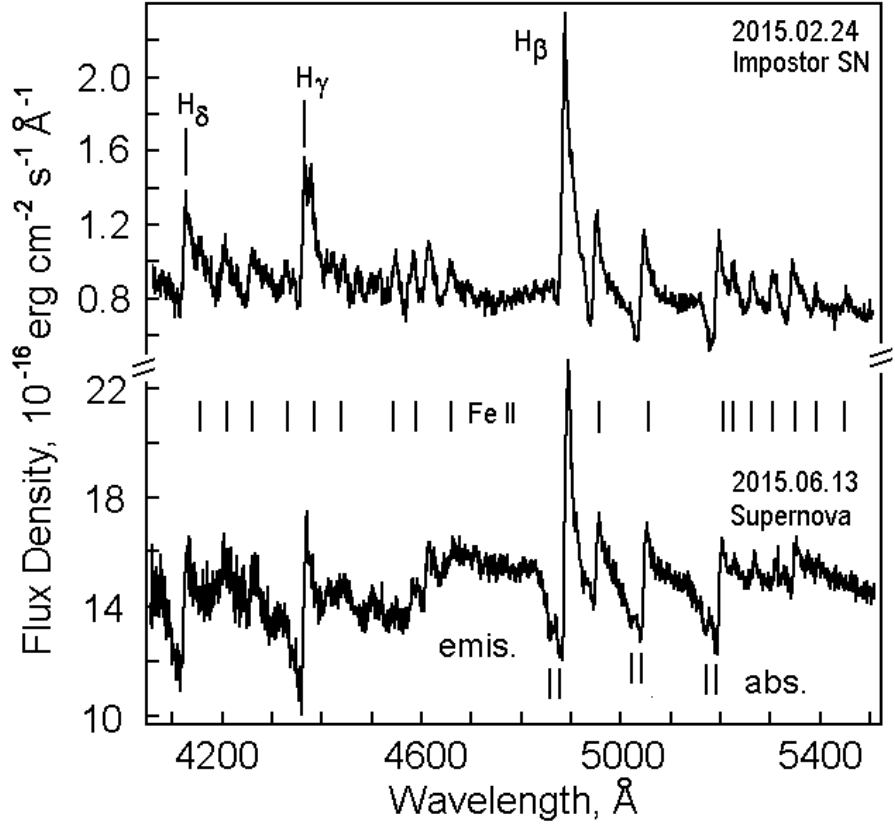}
\caption{
Spectra of SN~2015bh in the blue and green ranges at the impostor stage 2015a (top)
and supernova stage 2015b in 22 days after the brightness maximum (bottom).
$emis.$ is a high-velocity emission component in the H$_\beta$ line.
The bottom spectrum shows double narrow components of the circumstellar medium
($abs.$) formed before 2015b. 
}
\label{figure4}
\end{figure*}

\begin{figure*}
\includegraphics[scale=0.7]{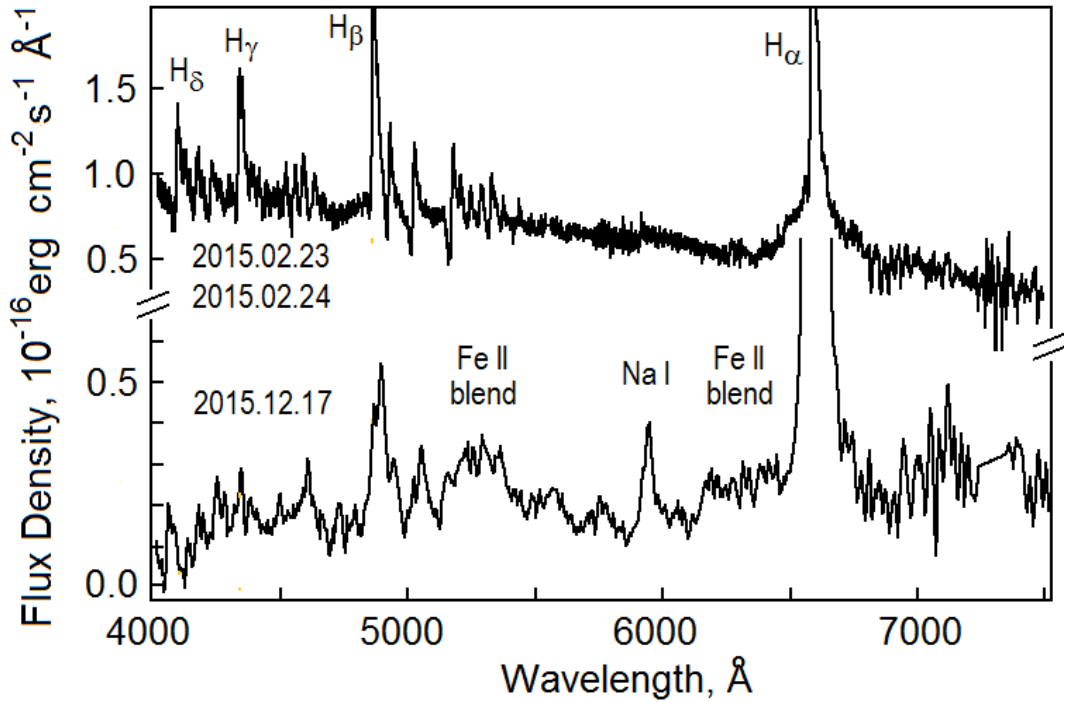}
\caption{Full spectra of SN~2015bh at the phase 2015a and 2015b 
at the secondary brightness decay. 
}
\label{figure5}
\end{figure*}

Fig. ~\ref{figure4} shows the spectra of SN~2015bh in the blue and green range
at the phase 2015a (top) and at the phase 2015b immediately after the brightness
maximum (bottom).
The total spectrum of the star at the phase 2015a composed of the spectra taken on
February 23 and 24, 2015 including the red range is shown in Fig. ~\ref{figure5} 
(top). Approximately, these spectra correspond to descriptions 
given by \citet{ERea15,UPLea15} and \citet{UPTea15}.

In the spectrum of February 23, 2015 at the phase 2015a, the H$_\alpha$ profile
consists of a narrow emission line with a weak absorption component shifted
to the blue side, both ones superimposed on a broad emission base.
The emission peak is at a velocity of 1900 \kms, or, taking into account the galaxy 
redshift, at --47 \kms. The width of the narrow emission component at the
half-maximum intensity level corrected for the instrument profile, $FWHM$ is 640 \kms.
This component has an emission hump in the red wing at a velocity of +870 \kms. 
The narrow absorption component is located at a velocity of --900 \kms,
symmetrically relative to the emission maximum. These velocities are measured
relative to the emission peak. The wide wings of H$_\alpha$ emission extend up
to 10000 \kms\ to each side from the peak. The H$_\beta$ line profile has the identical
structure, there is also a hump at a velocity of +870 \kms. In emission of this
line, one can trace only the short-wavelength wing extending up to --8500 \kms. 
The long-wavelength wing is distorted by strong Fe~II emission lines. 
Undoubtedly, the wide wings of Balmer lines are not related to motions of 
large mass of material, but they result from Thomson scattering of Balmer photons
by free electrons. The strongest Fe~II lines show the P~Cyg type profiles. 

In the blue spectrum of June 13, 2015 (the phase 2015b), a wide and deep
absorption component extending to the velocities --4000 \kms\ appeared in
the profiles of Balmer lines. In the depth of this component, a narrow absorption
detail at a velocity --900 \kms\ observed previously at the phase 2015a, remained,
and a new absorption detail at a velocity of --1900 \kms\ appeared (these details
were marked in Fig.~\ref{figure4}). 
Similar absorptions with double absorption details were observed in the
strong Fe~II lines at the phase 2015b. A considerable asymmetry of emission
components in the P~Cyg profiles of H$_\alpha$, H$_\beta$, strong emissions
Fe~II, their triangle shape are due to excess absorption in the blue wing
of the profile within the velocity range between --3300 and  --250 \kms.
(This fact can be set if superimpose the profile with its mirror reflection,
as was done, e.g., in Fig.~6 by \citet{SAvDea16} for the spectrum of the luminous
red nova NGC~4490-OT2011).
Such a structure of profiles is formed by dense optically-thick ejected shell,
its formation continued as early as at the phase 2015a.
The observation of hot continuum with the Balmer emissions, the lines He~II and He~I 
on May 16, 2015 by \citet{DBLea15} indicated the exit of shock wave 
at the brightness rise of the 2015b outburst. Because of this, during a month, 
the high-velocity absorption components appeared in the profiles, but the
component with a velocity of --900 \kms\ formed before the 2015 events,
remained in the line profiles. It is formed
in a medium where the shock wave has not reached yet. At the phase 2015b
a part of this medium got acceleration by the light pressure of the outburst,
and the second narrow absorption component at a velocity of --1900 \kms\ arose.
The interesting observational fact is that the velocity of the narrow absorption
component \hbox{--900 \kms} in the blue wing of the Balmer line profiles is
equal but oppositely directed to
the velocity of the emission component (hump) in the red wing. This
may mean that the initial medium was formed by a bipolar wind which was outflowing 
at a big space angle. A part of this conic outflow directed to the observer 
absorbs the light of the optically thick shell, and the opposite part is seen
in emission. Besides, these narrow components show that the primary circumstellar
medium was expanding to considerably larger distances than photosphere of the
ejected shell forming the P~Cyg type profiles.

Strong Fe~II lines are a typical feature of the SN~2015bh spectra at both 
phases, 2015a and 2015b. With these lines, the spectra resemble ones
of \citet{w12} Fe~II type classical novae more, than those of supernovae. 
Classical novae are cataclysmic or symbiotic systems comprising a far-evolved
star (a white dwarf) and a normal companion (a red dwarf or a red giant) which
is an accretion donor.
The outburst of a nova is a thermonuclear runaway of a hydrogen envelope 
accumulated on the surface of a white dwarf due to accretion. Of course, 
the SN~2015bh explosion scale exceeds the scale of classical novae. 
Its luminosity at the phase 2015a exceeds the maximum luminosity of
brightest classical novae 100 times, and at the phase 2015b does 1600
times. The formal resemblance between spectra of novae and SN~2015bh demands
a detailed study, because   
\citet{UPLea15,DBLea15} note presence of the He~I and even He~II lines,
which are not observed in Fe~II type classical novae except rare cases of
hybrid novae. Classical novae of another type, the Williams He/N, have the
He~I and He~II lines, and this is the main sign of the He/N type, but the iron
lines are not found in their spectra. 
The strongest He~I lines in the optical spectrum are expected at the wavelengths 
4121, 4143, 4388, 4471, 4713, 4922, 5016, 5876 and 6678~\AA.
Resolution of our spectra is sufficient to determine that the He~I components
are absent in possible blends with strong emissions of Fe~II 4922 and 5018~\AA\
(the multiplet 42), 
and in the blend with Na~I 5889/5896~\AA\ at the phase 2015b. The emission
peaks at 4922 and 5018~\AA\ coincide with other Fe~II lines in radial velocity. 
Other He~I lines were also not detected. There is a faint detail in the
region of He~I 4471 \AA\ in the spectrum of June 13, 2015 on the 22 day after the
maximum at the 2015b phase, the identification of which with the helium line
is doubtful. Other He~I lines including components of the Fe~II blends mentioned
above are not detectable. Besides two mentioned in our spectra,
Fe~II emissions are reliably identified at
the following wavelengths in Angstroms (the multiplet number is in brackets): 
4179(28), 4233(27), 4303(27), 4352(27), 4417(27), 4523(38), 4556(37), 
5169(42), 5198(49), 5235(49), the blend 5276(49)+5284(41), 
5317(48,49), 5363(48) and 5425 (48,49).
All these lines are available in the list of emission lines of classical novae in
optical range (Table~2, \citet{w12}).
Besides these lines, we have found two lines at 4590 and 4635 \AA\ which are absent
in this list. Both emissions are strong in our spectra of the Williams Fe~II
type classical nova V496 Sct obtained on November 10 and 11, 2009. 
According to Fig.~1.10 by \citet{HM12}, both lines are available in the 
$\eta$ Car spectrum where they are identified as the blends Fe~II and Cr~II. 
At the phase 2015a, the absorption component of strongest lines Fe~II 4924, 5018
and 5169~\AA\ (multiplet 42) reached the velocity --1400 km/s. This is a typical
velocity of an ejected shell in classical Fe~II type novae.
The half-width of the emission component was $\sim$940 \kms.  
At the phase 2015b, the absorption component of the Fe~II-line P~Cyg profiles
became stronger as well as in the Balmer lines, and the limiting velocities increased
up to --4000 \kms. Such velocities of envelope ejection exceed velocities of fast
high-luminosity Fe~II type classical novae, but they are small for SNe. At that,
the half-width of the emission component remained identical to that at the phase
2015a, 940 \kms. 
 
In June 13, 2015 spectrum in the SN phase, we observed a very wide emission
extending to the blue side of the H$_\beta$ line to 18000 km/s with the maximum 
flux density at $\lambda$ $\sim$4640~\AA\ (it is marked as {\it emis.} in 
Fig.~\ref{figure4} below). The red part of this emission is overlaid
by the absorption component of the H$_\beta$ line. Unfortunately, the emission
is seen only in one spectrum, and we have no observations
in the H$_\alpha$ line near the maximum brightness, so, it needs confirmation 
by independent observations. So far, the emission of so high-velocity outflow
is the only available spectroscopic evidence that the event 2015b is a supernova.
It is possible that in the spectrum of May 16, 2015 by \citet{DBLea15}
(6 days before maximum) a fragment of this emission was identified 
with He~II 4686 \AA. Besides, in description of that spectrum, there are no
mentions about the strong Fe~II lines, which are seen in all our spectra. 
As a rule, the lines He~II and Fe~II are not observed in spectra simultaneously 
because of difference between ionization potentials at which these lines form,
and, naturally, one can see either ones or the others depending on 
temperature of the excitation source. 

In a case of a radially symmetric outflow of material at the velocity of 18000 \kms,
the absorbing and emitting envelope components moving with
the velocities less than 4000 \kms\ would be swept out by this material, or 
the high-velocity gas would underwent a fast deceleration. Nevertheless,
a so unusual symbiosis of properties is observed in the spectrum. This symbiosis
can be explained only by the axial asymmetry of outflow at the SN explosion.
That is to say, the high-velocity gas was ejected along the polar axis whereas  
the slower outflow was occurring in equatorial directions. It is not excluded
that in the equatorial plane there already was a dense gas disk which was formed 
at an earlier stage of evolution or at the phase 2015a and it prevented from 
distribution of high-velocity gas in the equatorial plane. 
For example, spectropolarimetry of SN~2009ip by \citet{MGSea14} in the 
analogous phases 2012a and 2012b during its outburst in 2012 has shown 
that the outflows in these phases were strongly non-spherical, and the angles of
polarization plane were orthogonal. In opinion by \citet{MGSea14},
in the case of 2009ip, the ejecta generated at the explosion of 2012a, 
in the phase 2012b were colliding in the equatorial plane with a flattened  
(toroidal) structure. Evidences of bipolar ejections at LBV explosions are not
rare (this is seen in the Balmer line profiles, e.g., by \citet{BGVK14},
or in a bipolar structure of the "Homunculus"\ nebula around the galactic
LBV-type star $\eta$ Car).
At the case of SN~2015bh, probably, there was a "shot"\ of high-velocity gas 
in the observer direction related with the phase 2015b, whereas an opposite
ejection was overlapped by a body of a thick equatorial disk or a toroidal
structure at a relatively small inclination angle between this structure and
the line of sight.  

\begin{figure*}
\includegraphics[scale=0.9]{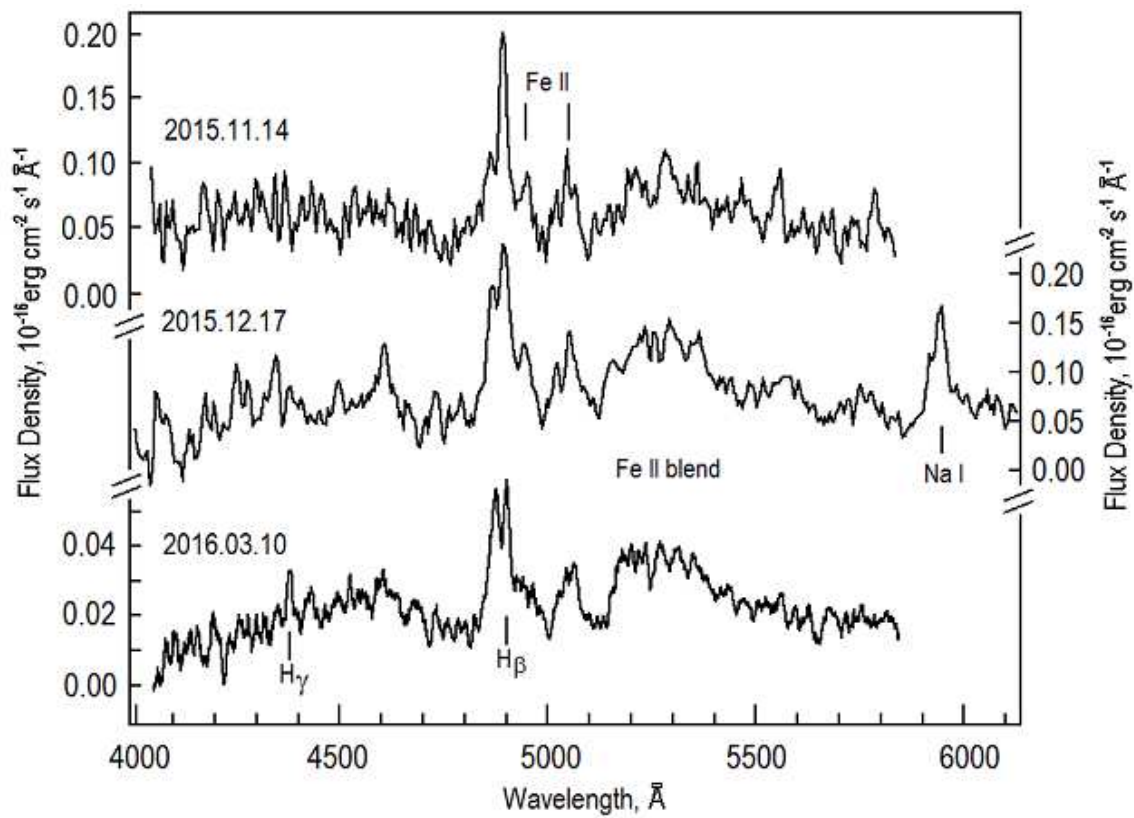}
\caption{
Spectral changes of SN~2015bh at the secondary brightness decay.
}
\label{figure6}
\end{figure*}

The spectra of the object in the blue and green ranges obtained at the
phase 2015b after it appeared from behind the Sun, i.e. at the secondary decay
of brightness (176--374 days from the brightness maximum) are shown in  
Fig.~\ref{figure6}, and the total spectrum in day 209 -- in Fig.~\ref{figure5}
(bottom). In blue spectra, strengthening of the short-wave emission component in
H$_\beta$ line profiles was seen, this component being located at the place of
the absorption component early observed near the light maximum. The same
emission components grew at the same part of Fe~II profiles
what led to appearance of an extended blend in the region $\lambda$4400 --
4700 \AA\ and 5150 -- 5450 \AA.
If there were no strong hydrogen lines, such a spectrum would be similar
to that of SN~Ic 1994I, 2003jd, 2010gx, 2004aw (see Fig.~4 by \citet{PSBea10}).

\begin{figure*}
\includegraphics[scale=0.9]{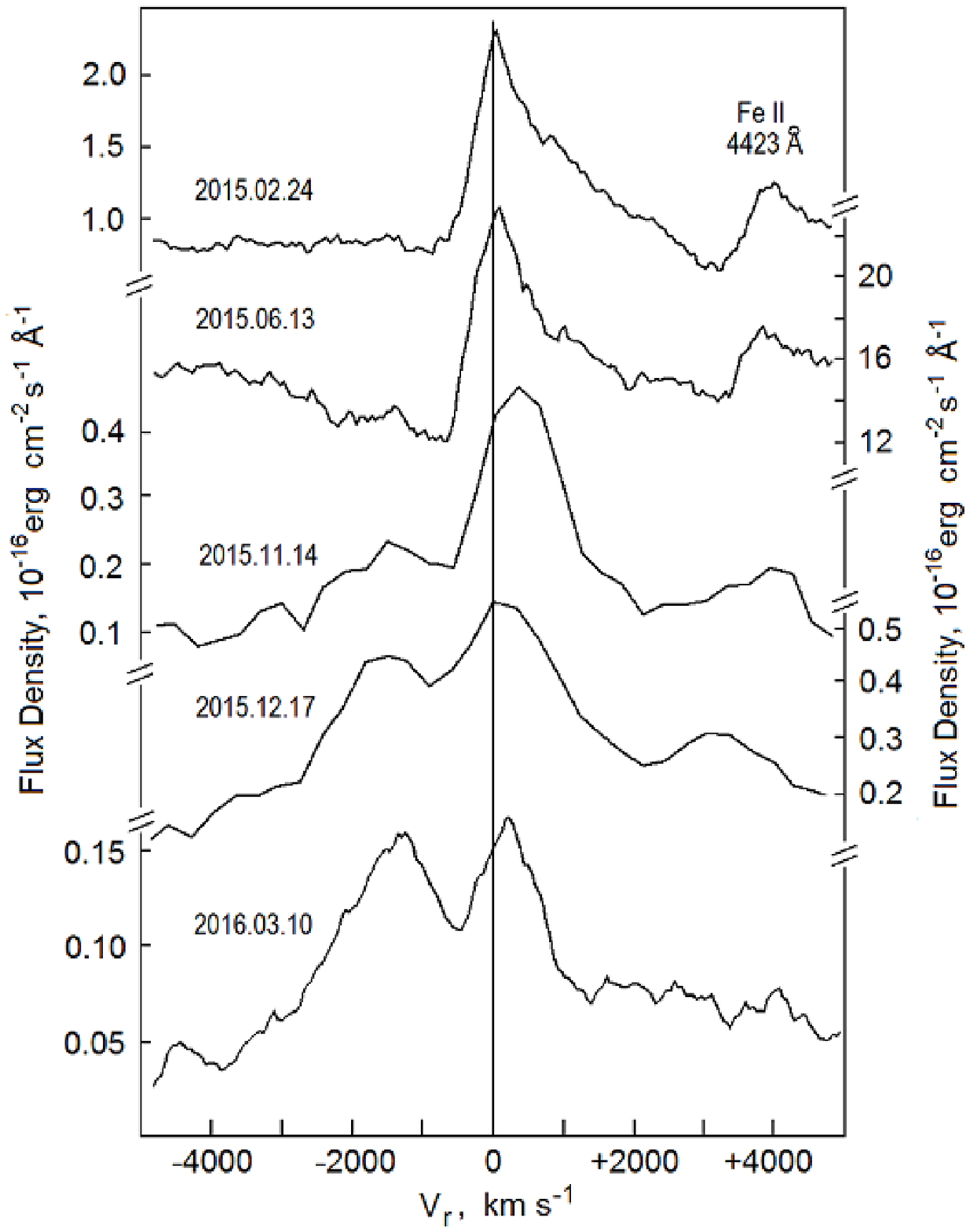}
\caption{
Profiles of the spectral line H$_\beta$ and their changes with time (from top to bottom). 
}
\label{figure7}
\end{figure*}

\begin{figure*}
\includegraphics[scale=0.8]{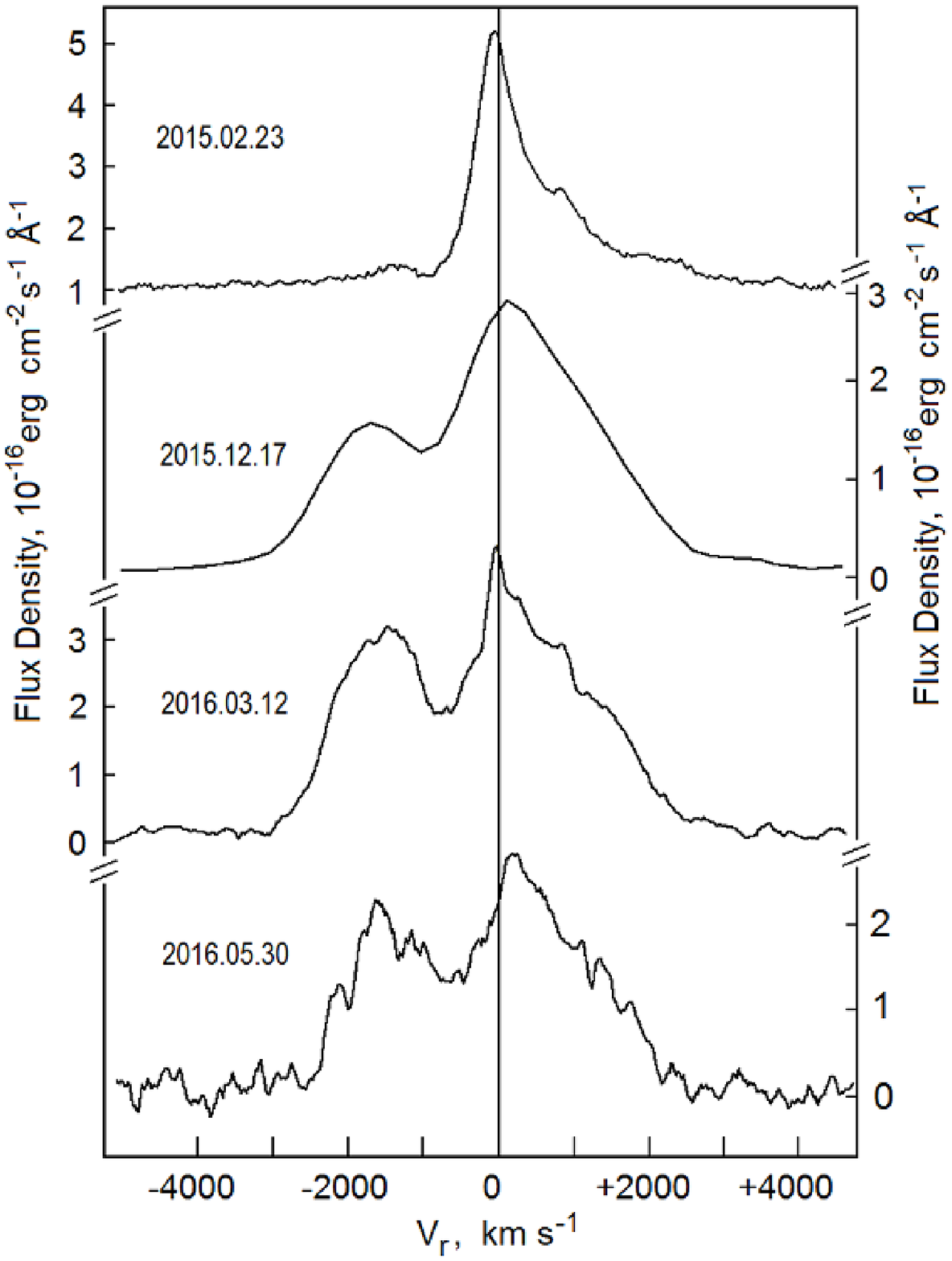}
\caption{
Profiles of the spectral line H$_\alpha$ and their changes with time (from top to bottom). 
}
\label{figure8}
\end{figure*}

Changes in the H$_\beta$ profiles are shown in Fig.~\ref{figure7}. 
Note that the December 2015 spectrum was obtained with
the low spectral resolution. In June 13, 2015 spectrum, the absorption
component extended to the short-wave side from --400 to --4000 \kms.
This velocity range in the profile was gradually filled with emission, and
on March 10, 2016
we already see in the spectrum a strong emission with a peak at --1300 \kms. 
In the H$_\alpha$ profiles shown in Fig.~\ref{figure8}, the similar effect
is seen
(except a profile with the absorption component of June 13, 2015; at that time
the red-range spectrum was not obtained). 
In the May 30, 2016 spectrum, the continuum was already lower than the noise level,
but the H$_\alpha$ profile remained emissive and kept the double shape. 

The identical behavior was described at late stages of light curve of the luminous
red nova NGC~4490-OT2011 \citep{SAvDea16}. Some difference from the event 2015bh
was a smaller range of the absorption component velocities, from 0 to --650 \kms,
so, the blue emission wing of the H$_\alpha$ profile was covered with the absorption
only partly, and an emission hump was observed at --650 \kms. The maximum absorption
depth in this profile, and then the maximum strength of emission were observed
at the velocity of --280 \kms. These velocities are 4--5 times lower than those
ones of SN 2015bh.
\citet{SAvDea16} give an entirely plausible explanation of such evolution 
of the profile: "a massive shell was ejected at these speeds and was initially seen
in absorption, but as the underlying photosphere cooled (as indicated by the
spectral evolution during the decline from peak) and as the optical depth dropped,
the same dense shell is seen in H$_\alpha$ emission. This likely indicates ongoing
shock heating of this ejected shell". To explain transition
from absorption to emission in 2015bh one should add a factor of
fast expansion of the ejected shell ($v_r$ up to 4000 \kms) causing it to become
optically thin quickly. \citet{SAvDea16} interpreted the event NGC~4490-OT2011
as a luminous red nova, the merger in a massive system.
As a rule, the expanding envelopes of luminous
red novae do not transit to the optically thin phase at all, and if they do,
it turns out to be cold unionized and even molecular gas, as in the case of
V4332 Sgr. Note that the NGC~4490-OT2011 light curve had two maxima, and in the
second outburst (with a cold-star spectrum) it achieved a record absolute
magnitude  --14\fm\ At merging of cores of massive stars, a shock wave forms,
and when it comes out on the surface, the first brightness maximum is
observed (an example is LRN~2015/M~101, also a massive system studied by
\citet{GBSea16}). 

\section{DISCUSSION OF RESULTS}

The relation between SNe impostors and LBVs is established quite reliably. Some
SNe~IIn are also obviously related with LBVs. Specific evidences of these
relations were considered in detail by \citet{TPSea16} and in studies cited there. 
However, only four impostors are known which eruptions led to appearance of SNe~IIn.
Besides the above-mentioned event 2009ip, these are 2010mc \citep{SMP14}, 
2011ht \citep{FMKea13}, and LSQ13zm \citep{TPSea16}. 
SN~2015bh demonstrates a certain similarity with these objects, but there are
differences both in light curves and in spectra. The nature of LBV explosions,
which lead to "detonation"\ of more large-scale explosions of SNe still remains
unclear, although there are several hypotheses explaining this phenomenon.

(1) PPI -- "pulsation pair instability", related to massive formation of 
electron-positron pairs in cores of massive hot stars, which results in a
reduction of radiation pressure, a partial collapse of a star and a thermonuclear
explosion. 
The thermonuclear explosion stops the collapse and gives rise to ejection of
the envelope. In this case, the repeated outburst can be explained by interaction
between ejecta and dense circumstellar medium \citep{WBH07}. The detailed review
of final stages of evolution and mechanisms of SNe in massive
stars is given by \citet{H12}. It was established that the thermal energy
released at the collision of rapidly moving ejecta with the dense 
circumstellar medium may be higher than that of the core collapse.
Note that the PPI mechanism is valid for stars with initial mass within 
80 -- 280 M$_\odot$, and it can generate repeated eruptions of an impostor,
similar to the case of SN~2009ip, and, eventually, leads either to the core
collapse  and formation of a black hole, or to a full destruction of the star
(at the photodisintegration of nickel for a mass within the range
130 -- 280 M$_\odot$).

(2) Core collapse of a faint supernova and interaction of ejecta with a medium 
\citep{MSF13}.

(3) Repeated outbursts related with approaching of components at elliptical
orbits with the interaction in a binary system, and a final outburst
at the merger or direct collision (this hypothesis was considered for SN 2009ip,
in which several outbursts of an impostor were observed before the SN explosion
\citep{SK13}).

(4) Last explosion of LBV (a SN impostor) followed by the core collapse 
(SN~IIn) \citep{TPSea16}.

These hypotheses were considered in detail by \citep{TPSea16}. We think one more 
scenario is possible. 

(5) Merger of components of a massive system with formation of a common
envelope broken by the core collapse of one of stars being at a later evolutionary
stage because of an excess of the critical mass (a failed luminous red nova in a 
massive binary system). In such a case, the process of the merger of stars does
not need to end by merging of their cores, and then the SN explosion leads to
a binary system containing an optical remnant of a companion (the accretion donor)
+ a compact component. 

The scenario determines if a SN is a final stage of evolution of a massive star
finished by a compact object, or it is a single event in an LBV history.
This question is to be solved by observations of remnants of similar double
explosions. Some events were discovered as SN~IIn (e.g., LSQ13zm), but then their
relation with an impostor might be detected from archive data. Are all type IIn
SNe related to previous LBV explosions? This question also requires special studies.

Now let us consider how each hypothesis agrees with observations of SN 2015bh.

(1) The PPI hypothesis. Can the spectral similarities of SN 2015bh with classical
Fe~II type novae we found in this outburst be interpreted in favor of this
hypothesis? Classical novae can be attributed based on their spectra to one of 
two types, He/N and Fe~II, or they can be "hybrid"\ with transition from 
Fe~II to He/N at an outburst. \citet{w12} explains this difference by
the origin of gas ejected in the explosion. The He/N type novae 
have in their spectra very wide emission lines (a high velocity of ejection) 
of elements higher ionization degree with the rectangular profiles and almost
always without absorption components. This means that the gas is ejected at a 
thermonuclear runaway just from the white dwarf. Spectra of Fe~II type novae
at the initial decay of light curve are characterized by numerous Fe~II lines
of low ionization degree which were excited by atomic collisions, though 
in the far red range there are lines of CNO elements which are excited at 
recombination and fluorescence scattering. Profiles of these lines are
rounded, more narrow and often have P~Cyg-type absorption features of optically
thick expanding gas. The intensities of the Fe~II emissions decrease relatively
slowly as the brightness decreases, and occasionally increase again in the
episodes of "secondary maxima"\ which happen during several months after the
beginning of the outburst.
In the brightness maximum and in secondary maxima at the decay of brightness 
the emitting gas is colder than in other phases.
At the early decay of brightness, Fe~II novae display narrow absorptions
of heavy elements of the iron peak. This means that the spectrum forms in the
gas shell with the element composition close to the solar one. \citet{w12}
assumes that this gas arises from the secondary component, a normal star 
rather than from a white dwarf. Probably, it is not even mixed with the explosion
ejecta, and saturated with heavy elements generated in the process of evolution
of the secondary star. Hydrodynamic calculations show that the radiation and
the material thrown off the white dwarf stimulate the secondary component to lose
the mass through the Lagrange points L1 and L3, and the shock from the orbital
motion and accretion disk forms a circumstellar shell. At the same time,
it was established that hybrid novae develop only from the type Fe~IIb novae
(with broad lines in spectra), but not from Fe~IIn ones (with narrow lines). 

The arguments against the PPI hypothesis are as follows:

(a) The PPI hypothesis does not suppose that there is a component in a binary
system. However, from the observed spectrum of SN 2015bh it follows that
the composition of elements in the ejected shell and stellar wind at the outburst
is close to solar one or only slightly evolved. As in the classical Fe~II type
novae, this material can originate in accretion from the secondary companion.  

(b) Within the context of this hypothesis, the PPI event and thermonuclear
explosion explain the phase 2015a, and the transformation of kinetic energy of
ejecta into emission at interaction with circumstellar medium explains the 2015b
phase. Photometric and spectral signs of SN at the 2015b phase noted by us
previously, contradict the PPI hypothesis. 

(c) The time of wave propagation inside the envelope released at the 2015a phase
can be estimated as an interval between the end of the 2015a phase (the beginning
of brightness decay at the moment JD~2457138, mag = 18\fm6 according to 
observations by Hoverton) and the outburst peak at the 2015b phase
JD 2457165, $V$ = 15\fm35. This is equal to 27 days. Such a term is too large
for propagation of dynamic changes, but it is typical for transfer of heat and
radiation in massive envelopes (in the luminous red novae).  

(2) Core collapse of a low-mass star, a faint supernova and interaction
between the ejected shell and circumstellar medium. At that, the 2015a phase
is interpreted as an SN~IIP outburst, and the 2015b phase as interaction between
the ejecta and environment. The examples of faint core-collapse SNe~IIP
reaching the absolute magnitude --14$^m$ in the maximum brightness, 1999br and
2010id are adduced by \citet{TPSea16}. For SN~2015bh there are no evidences of 
a supernova outburst in the phase 2015a, although at the 2015b phase they are
available: the rate of the secondary decay in the light curve is the same as
the calculated $^{56}$Co decay rate, emission in the H$_\beta$ line
with the velocity up to 18000 km/s. These are arguments against the hypothesis 
about a faint SN. 

(3) Approaching and final merging of components, a massive evolved star
of mass 60-100 M$_\odot$ in a system with a star of mass 12--50 M$_\odot$ moving
on an elliptic orbit. Paper by \citet{SK13} gives weighty arguments in favor of
this hypothesis for SN~2009ip. We have no such arguments for the event 2015bh --
repeated outbursts of the impostor or humps at the brightness decay at the burst
of 2015b. Although \citet{TPL16} detected "short-term variability at least during
21 years"\ before the event of 2015. The merger of components even for massive
stars leads to the phenomenon of a red nova, a cold explosion, but we have no
spectral evidence of a red nova. 

(4) The last explosion of LBV (the SN impostor) and a core collapse (SN~II).  
The PPI event which provoked collapse of a massive core of LBV at a late stage 
of its evolution may be considered as the cause of explosion. The late stage of 
LBV evolution can be testified by unusually strong lines of iron. Iron could get
to the star envelope due to the mixing. This is one of the promising hypotheses. 
In that case the explosion remnant will be a single black hole.

(5) Collapse of a core of a massive evolved LBV star in the system in a merger
with a massive companion caused by exceeding critical mass of the core (a failed
red nova in a massive binary system). E.g., in the galactic system \hbox{$\eta$ Car,}
in the orbit around LBV there is an O-class star of early or middle subclass
with a mass within the range \hbox{30--60 M$_\odot$,} which passed a less
evolutionary track than the main more massive component \citep{CI12}. Before the
2015b outburst, a gradual rise of the stellar brightness was observed which finished
by a decline by 1$^m$. The 2015a phase is usually called a SN impostor, but
impostor light curves do not show gradual increase of brightness during outbursts. 
On the contrary, they show a decline typical for SNe (maybe, except SN~1954J).
Such a brightness rise is typical for luminous red novae, and it is related with
the formation of a common envelope at merger. The similar effect was observed in
the luminous red novae V4332 Sgr, V1309 Sco, LRN~2015/M31 (M31N 2015-01a) and 
LRN~2015/M~101 (but it was absent in V838 Mon). The brightness decline by 1$^m$
before the luminous red nova outburst was observed in the case of
V1309 Sco. \citet{BGVZ14} explain this decline by the merger of star cores and
a slow shock from the star center which leads to transfer of the envelope
in the expansion mode in the regime close to the adiabatic one. 
The term "slow shock"\ was first used by \citet{MWT99} to explain
the phenomenon of luminous red nova for V4332 Sgr.

At the merger of star cores in the massive red novae NGC~4490-OT2011
\citep{SAvDea16}  
and LRN~2015/M~101 \citep{GBSea16}, a shock wave formed, and, after it came out
to the surface, the envelope passed to the mode of adiabatic expansion, which is 
accompanied by a deep weakening of brightness before the secondary maximum.
Up to now, no spectra were obtained for the stage of common envelope formation in
a luminous red nova, and, most probably, they are not similar to those of classical
type Fe~II novae. Such a spectrum of SN~2015bh at the 2015a phase can be explained
by accretion of hydrogen-rich material from the companion onto the evolved star
(LBV) at the phase of a common envelope formation, development of thermonuclear
burning in a layer and release of additional energy by the accretion. 
At a high luminosity exceeding the Eddington limit, the mass outflow occurs with
the typical rate for classical novae in outburst.
At the merger of stars, the angular momentum is carried off with the material through 
the Lagrange point L1, which results in formation of a disk surrounding the system.
Formation of a thick disk around the massive star inside a common
envelope can also be expected. In the moment when the critical mass was exceeded,
the core collapse of evolved star occurred, and under such conditions, the
explosion was asymmetric, with high-velocity gas outflow in the direction of
axis of the inner disk.
In the specific case of SN~2015bh, we observe the luminosity decline at the end 
of the 2015a phase, i.e. the slow shock already occurred, and the adiabatic
expansion of envelope started. However, the core collapse stopped the adiabatic
expansion of the envelope and interrupted the scenario of a luminous red nova.
That is why the interrupted scenario can be called "a failed luminous red nova".
Propagation of the shock wave inside the common envelope and its coming out 
to the surface can explain the light curve at the 2015b phase with its high
luminosity and a sharp peak in maximum, along with a high temperature at the
2015b outburst decay (in contrast to luminous red novae), and transition of the
envelope to the optically thin state. 
The abundance of hydrogen in the spectrum of 2015bh, the origin of which can be
in the envelope of the less evolved companion, can argue for hypothesis (5). 
This hypothesis predicts a remnant being a binary system with a relativistic
component, if the collapse occurred earlier than the total merger of the cores.

The latter hypothesis explains the events observed in the light curve 
and in the star spectrum in details. However, some components of the scenario 
(the high-velocity ejection in the H$_\beta$ profile, the luminosity decline at
the transition of the envelope to the mode of adiabatic expansion) are based
on single observations and need a confirmation. The question of whether 
the already expanding envelope can transfer to the adiabatic mode due to
a shock from within the star center remains open. The dynamical simulations are
required. On April 7, 2015 Zhijang Xu and Xing Gao observed a short burst which
may be related with the exit of a primary, weaker shockwave. This observation
should be verified. 
The question of whether the 2015bh event is a final episode in the evolution
of a massive star is not yet solved, too. It will be answered by the study of
the explosion remnant with large telescopes at a level of stellar magnitudes
larger than 25\fm

\section{CONCLUSIONS}

Spectra of the optical transient 2015bh/NGC~2770 in the phases of SN impostor
and real SN obtained with BTA/SCORPIO are unusual, and they are similar to 
spectra of Williams Fe~II type classical novae. From spectral data it was
established that in the impostor phase (2015a), an optically thick expanding shell
formed which was passed by a shock wave at the SN phase (2015b). Eventually,
the shell was accelerated and transferred to the optically thin phase.
Besides, we have revealed the absorption components in spectral line profiles
related to the extended circumstellar medium which partly accelerated by
radiation pressure of SN explosion. Near brightness maximum, we have detected
the emission component of high-speed, 18000 \kms\ ejecta in the H$_\beta$ line
profile along with slower moving, up to 4000 \kms\ absorption components, what
suggests the asymmetric eruption.
The light curve resembles the SN IIL ones with the rate of the secondary decline
corresponding to the rates caused by the radioactive decay of
$^{56}$Co isotope. In maximum, the SN reached the absolute magnitude
$M_V$ = --18\fm14 $\pm$0\fm30.

The SN~2015bh progenitor is a luminous blue variable star (LBV) with the strong
emission H$\alpha$, one of the brightest stars of the galaxy NGC~2770.

We consider that the most probable hypothesis to explain the SN~2015bh event
is a core collapse of a more massive far-evolved star in a binary system breaking
its merging with a less massive companion (a failed luminous red nova in a massive
binary system). Arguments for and against other hypotheses were also considered.
The core collapse of a single massive star induced by the PPI event is also
possible.\\

\section{Acknowledgments}

In the paper we used the Sloan Digital Sky Survey database, the NASA/IPAC
Extragalactic Database (NED), the Hubble Space Telescope archive,
the hosting site of images Flickr.com, the databases of supernovae
by Rochester Astronomy and Sternberg Astronomical Institute of the Moscow
University.
The spectral and photometric observations performed in the Special Astrophysical
Observatory of the Russian Academy of Sciences, their processing and analysis
were financed by the Russian Science Foundation (RSF) by the Grant No.14--50--00043.
The operation of the Russian 6 meter telescope BTA is financially supported by
the Ministry of Education and Science of the Russian Federation.
V.P.G., E.A.B. and A.F.V. thank the Russian Foundation for Basic Research (RFBR)
for the financial support of this work with the Grant 14--02--00759.
A.V.Zh. thanks RFBR for the support by the Grant 16--02--0758. The research by 
D.Yu.Ts. was partially supported by the RSF Grant 16--12--10519. The work by 
I.M.V. was partially supported by a Grant of the Slovak Information Agency SAIA.



\newpage

\newpage

\newpage

\newpage

\newpage

\newpage

\begin{table*}
\caption{Photometry of SN~2015bh in NGC 2770}\label{table1}

\begin{tabular}{l|c|c|c|c|l}
\hline
JD hel.24...&  $B$      &  $V$        &  $R_C$     & $I_C$ 	& \ \ remark\\
\hline
57076.491 & 19.65 (5)   & 19.26 (5)   & 18.71  (4)  & 18.60  (5)  &\ \  6m \\
57078.373 & 19.58 (8)   & 19.19 (5)   & 18.576 (40) & 18.44  (5)  &\ \  6m \\
57100.284 & 19.372 (50) & 18.841 (30) & 18.446 (20) &      -      &\ \  SO \\
57164.255 & 15.579 (40) & 15.376 (20) & 15.214 (20) & 15.167 (40) &\ \  SO \\
57164.266 & 15.581 (40) & 15.368 (20) & 15.214 (20) &      -      &\ \  SO \\
57165.248 & 15.566 (40) & 15.347 (20) & 15.205 (20) & 15.101 (40) &\ \  SO$^1)$\\
57165.338 & 15.54 (5)   & 15.35 (3)   & 15.18 (4)   &      -      &\ \  IW \\
57166.320 &      -      & 15.272 (20) & 15.136 (20) &      -      &\ \  SO \\
57167.28  & 15.572 (60) & 15.362 (50) & 15.265 (40) & 14.978 (50) &\ \  IW \\
57168.29  & 15.611 (70) & 15.387 (50) & 15.193 (50) & 15.007 (50) &\ \  IW \\
57169.32  &     -       &      -      & 15.24 (6)   &      -      &\ \  A2 \\
57174.27  & 15.906 (70) & 15.558 (50) & 15.399 (50) & 15.232 (50) &\ \  IV \\
57175.28  & 16.022 (80) & 15.618 (20) & 15.453 (50) & 15.224 (50) &\ \  IV \\
57175.30  & 16.002 (70) & 15.645 (20) & 15.423 (50) & 15.213 (50) &\ \  IV \\
57176.36  & 16.137 (70) & 15.714 (50) & 15.444 (50) & 15.249 (50) &\ \  IV \\
57177.34  & 16.106 (50) & 15.665 (30) & 15.415 (50) & 15.09 (1)   &\ \  IV \\
57179.31  & 16.084 (70) & 15.768 (50) & 15.497 (50) & 15.232 (50) &\ \  IV \\
57180.3   & 16.213 (60) & 15.766 (50) & 15.450 (40) & 15.07 (1)   &\ \  A5 \\
57181.3   & 16.310 (60) & 15.863 (50) & 15.547 (50) & 15.281 (50) &\ \  IV \\
57182.3   & 16.252 (60) & 15.883 (50) & 15.587 (40) & 15.340 (40) &\ \  A5 \\
57183.3   & 16.301 (60) & 15.868 (50) & 15.691 (40) & 15.366 (40) &\ \  A5 \\
57184.257 &      -      & 15.912 (20) & 15.676 (40) &      -      &\ \  SO \\
57184.27  & 16.420 (70) & 15.923 (20) & 15.639 (30) & 15.330 (40) &\ \  IV \\
57185.272 & 16.513 (30) & 15.931 (20) & 15.644 (20) & 15.374 (40) &\ \  SO$^2)$ \\
57186.29  & 16.541 (70) & 15.936 (20) & 15.690 (40) & 15.384 (40) &\ \  IV \\
57187.285 &      -      & 16.005 (30) &      -      &      -      &\ \  6m \\
57188.33  & 16.648 (70) & 16.090 (50) & 15.777 (40) & 15.487 (50) &\ \  IV \\
57190.3   & 16.726 (60) & 16.073 (50) & 15.805 (50) & 15.402 (40) &\ \  A5 \\
57276.55  &     -       &      -      & 18.31  (20) & 17.49  (15) &\ \  SO \\
57313.59  &     -       &      -      & 19.45  (11) &       -     &\ \  FL \\
57330.58  &     -       &      -      & 19.64  (10) &       -     &\ \  A5 \\
57331.58  &     -       &      -      &     -       & 18.71  (8)  &\ \  A5 \\
57332.58  &     -       & 20.33 (20)  &     -       &       -     &\ \  A5 \\
57340.45  &     -       &      -      & 19.77 (14)  &      -      &\ \  A5 \\
57340.54  & 21.37 (17)  & 20.67 (12)  & 19.57  (2)  &      -      &\ \  6m \\
57369.50  &     -       & 20.84 (22)  & 19.69  (3)  &      -      &\ \  SO \\
57373.53  & 21.48 (6)   & 20.85 (08)  & 19.71  (11) &      -      &\ \  6m \\
57426.483 &     -       & 21.60 (12)  & 20.33  (2)  & 20.27  (20) &\ \  SO \\
57456.468 & 22.16 (9)   & 21.95 (18)  & 20.48  (3)  &      -      &\ \  SO \\
57457.404 &     -       &      -      &     -       & 20.58  (11) &\ \  SO \\
57458.326 & 21.96 (12)  & 21.56 (12)  & 20.20  (5)  &      -      &\ \  6m \\
57512.311 &     -       & 21.96 (17)  & 20.47  (5)  &      -      &\ \  SO \\
57539.3   &     -       & 22.4 (9)    &     -       &      -      &\ \  6m \\
\hline
\multicolumn{6}{l}{\footnotesize {\bf Remarks:}}\\
\multicolumn{6}{l}{\footnotesize $^1$ $U$ = 14.75 (3)}\\
\multicolumn{6}{l}{\footnotesize $^2$ $U$ = 16.21 (5)}\\
\multicolumn{6}{l}{\footnotesize
6m -- the 6-m telescope BTA and the light reducer SCORPIO with the $BVR_CI_C$ filters }\\
\multicolumn{6}{l}{\footnotesize
\citep{Afanas05};}\\
\multicolumn{6}{l}{\footnotesize
A2 -- the 70-cm telescope AZT-2 of SAI in Moscow with the CCD Apogee Ap-7;}\\
\multicolumn{6}{l}{\footnotesize
A5 -- the 50-cm Maksutov meniscus  telescope of the Crimean station of MSU with the CCD}\\
\multicolumn{6}{l}{\footnotesize
\ \ \ \ \ \ Apogee Alta U8300;}\\
\multicolumn{6}{l}{\footnotesize
FL -- the 60-cm Zeiss telescope of the Crimean station of MSU and $UBVR_CR_JI_J$-}\\
\multicolumn{6}{l}{\footnotesize
\ \ \ \ \ \ photometer with the CCD camera FLI PL4022;}\\
\multicolumn{6}{l}{\footnotesize
IV -- the 1-m Zeiss telescope of Simeiz Observatory in Crimea with the CCD FLI PL09000;}\\
\multicolumn{6}{l}{\footnotesize
IW -- the 60-cm Zeiss telescope of Simeiz Observatory in Crimea with the CCD VersArray 512UV;}\\
\multicolumn{6}{l}{\footnotesize
SO -- the 1-m telescope of SAO RAS and the $UBVR_CI_C$-photometer with the CCD EEV 42-40.}
\end{tabular}
\end{table*}

\begin{table*}
\caption{Spectra of SN~2015bh in NGC 2770 taken with BTA/SCORPIO}\label{table2}

\begin{tabular}{c|c|r|c|c|c|c|c|c}
\hline
Date UT & JD hel.24...& $\Delta t$ & $\epsilon$, s & $\lambda$, \AA & R, \AA & Grism & $\Delta v_r$, \kms &\ S/N \\
\hline
2015.02.23  & 57076.5160 & -88 & 3600 & 5751 -- 7501 &  5.5 & VPHG1200R & -11.7 & 22 \\
2015.02.24  & 57078.3933 & -87 & 3600 & 4053 -- 5847 &  5.0 & VPHG1200G & -12.7 & 15 \\
2015.06.13  & 57187.2982 & +22 & 1800 & 4056 -- 5850 &  5.0 & VPHG1200G & -21.2 & 30 \\
2015.11.14  & 57340.5885 &+176 & 3600 & 4048 -- 5846 &  5.0 & VPHG1200G & +28.4 &  3 \\
2015.12.17  & 57373.5602 &+209 & 4800 & 3740 -- 7877 & 14.6 & VPHG550G  & +20.5 &  2 \\
2016.03.10  & 57458.2293 &+293 & 3600 & 4043 -- 5846 &  5.5 & VPHG1200G & -18.3 &  2 \\
2016.03.12  & 57460.3973 &+295 & 2400 & 6024 -- 7096 &  3.5 & VPHG1800R & -19.4 &  4 \\
2016.05.30  & 57539.3346 &+374 & 2400 & 6025 -- 7098 &  3.0 & VPHG1800R & -24.9 &\ \  $^1)$ \\
\hline
\multicolumn{9}{l}{\footnotesize $^1$ Spectral continuum is lower than the noise level. }\\
\end{tabular}
\end{table*}

\bsp	
\label{lastpage}
\end{document}